%% file: CryptoLotteries_v3.0.tex
\documentclass[11pt,a4paper]{article} 
\usepackage{amsmath}     
\usepackage{amssymb}
\usepackage{amsthm}
\usepackage{eurosym}
\usepackage{setspace}
\usepackage{url}

\usepackage{natbib}

\usepackage{sectsty}           
    \sectionfont{\large}       

\usepackage{color}

\usepackage[text]{diff} 

\usepackage[dvips]{graphics}    
\usepackage{caption}
\usepackage{float}
\usepackage{subfig}             

       \usepackage[a4paper]{anysize} 
       \marginsize{2.5cm}{2.5cm}{2.5cm}{.8cm}

\theoremstyle{plain}

\theoremstyle{definition}

\theoremstyle{remark}

\begin{document}

\linespread{1}\small\normalsize
\title{\textsc{\Large{Risks and Transaction Costs of Distributed-Ledger Fintech: Boundary Effects and Consequences
}}}

\author{Kim Kaivanto%
\thanks{Corresponding author: tel +44(0)1524594030; fax +44(0)1524594244; e-mail k.kaivanto@lancaster.ac.uk}
$^\ddag$
~and Daniel Prince$^\S$\\
$^\ddag$Department of Economics, Lancaster University, Lancaster LA1 4YX, UK\\
$^\S$Security Lancaster, Lancaster University, Lancaster LA1 4WA, UK}

\date{\emph{this version:} \today}

\maketitle
\thispagestyle{empty} 

\begin{abstract}
\noindent
Fintech business models based on distributed ledgers -- and their smart-contract variants in particular -- offer the prospect of democratizing access to faster, anywhere-accessible, lower cost, reliable-and-secure high-quality financial services.
In addition to holding great, economically transformative promise, these business models pose new, little-studied risks and transaction costs.
However, these risks and transaction costs are not evident during the demonstration and testing phases of development, when adopters and users are drawn from the community of developers themselves, as well as from among non-programmer fintech evangelists.
Hence, when the new risks and transaction costs become manifest -- as the fintech business models are rolled out across the wider economy -- the consequences may also appear to be new and surprising.
The present study represents an effort to get ahead of these developments by delineating  risks and transaction costs inherent in distributed-ledger- and smart-contracts-based fintech business models.
The analysis focuses on code risk and moral-hazard risk, as well as on mixed-economy risks and the unintended consequences of replicating bricks-and-mortar-generation contract forms within the ultra-low transaction-cost environment of fintech.
\end{abstract}

\vfill
\smallskip
\noindent \emph{Keywords:}\/\;
distributed ledger, blockchain, smart contracts, transaction costs, code risks, asymmetric information, moral hazard, market failure, information impactedness, regulation, fintech service design

\newpage

\input{CryptoLotteries-body_v3.0}





\end{document}

%% file: CryptoLotteries-body_v3.0.tex
\section{Introduction}
New financial technologies (fintech) based on Distributed-Ledger (DL) technologies hold great promise, and not without reason.
DL business models and especially their Smart-Contract (SC) variants offer great advantages, seemingly in every area.
Some of these advantages flow directly from abandoning the hallmark characteristics of traditional business models.
These characteristics may be broadly categorised as follows.
\emph{Physical presence}: physical-location specific, carrying overhead and constraints of bricks-and-mortar presence, relying on droves of employees to implement process routines by hand.
\emph{Organizational structure}: carrying the overhead and resultant skewed focus associated with the management functions needed to supervise these employees,
business-hours limitations and office-culture based processing and decision-making practices.
\emph{Business IT systems}: legacy information systems and backward-compatibility constraints on new systems.
\emph{Payment and settlement systems}: legacy payment and remittance systems, with associated settlement systems.
These hallmark characteristics form a loose taxonomy of the features present in incumbent systems and organizations.
Importantly they also feature contextually adapted intrinsic risk-mitigation systems inherent in their provision, some of which are non-obvious and which may not be taken for granted in DL implementations.

Other advantages flow from the properties of DL technologies:
e.g.
speed and real-time data-driven algorithmic decision processes,
automation,
native data-driven and machine-learning platform compatibility,
native strong security,
immutability,\footnote{due to the underlying distributed-consensus protocol through which new blocks are added to the blockchain, fixing permanently transactions recorded in previous blocks}
provable fairness,
geographically unlimited `reach' leading to liquidity and diversification benefits,
and vastly reduced fixed, variable and transaction costs.

Although the advantages of DL-based business models may dominate overall, it would be a mistake to surmize that fintech business models dominate on \emph{every line item} of cost and risk.\footnote{The tree of cost and risk divides into branches, from branches into sub-branches, and so on, until eventually terminating in `leaves'. In some of those leaves, fintech business models do not dominate traditional business models on cost and risk.}
It is important -- for fintech firms themselves, for customers, and for regulators -- to recognize that while some risks and Transaction Costs (TCs) are drastically lowered, at the same time some other (new) risks and TCs are enhanced.
It is more accurate to think of DL-based fintech as having a different \emph{profile} of risks and TCs -- including some new risks and TCs which are absent or mitigated in incumbent business models.

The concern in moving to a DL-based business model, which replicates functions from the traditional business while discarding inefficient elements of its hallmark characteristics, is that those features which inherently mitigate risk are also elided.
In this situation not only are there the emergent new risks of moving to a fundamentally new technology, but also those risk which accumulate in the system due to a lack of comparable inherent risk mitigation as in the incumbent business models.

In this paper we examine and characterize the risks and TCs of DL fintech business models, particularly insofar as they differ from those of incumbent business models.
The transition from bricks-and-mortar business models to DL business models creates new classes of unanticipated consequences, primarily through the new categories of risks and TCs within a drastically changed profile of risks and TCs.
Surprisingly however, the drastic reduction in TCs is not a universally unalloyed advantage.
Those business activities that \emph{rely} heavily upon TCs in the bricks-and-mortar world may break down entirely when implemented in the ultra-low TC environment of DL SCs.

Risks, TCs, and potential for problems arise at the boundary between the DL and the external economy.
Whereas the DL is distinguished by hard security, distributed consensus, and non-manipulability, these properties do not extend beyond the DL to the goods, assets, services, and currencies of the external traditional economy.
Indeed the use of DL technology in the form of SCs on external-economy assets can raise the stakes involved in -- and therefore the incentives for -- attacks on the orderly operation of the  distributed-consensus algorithm to levels manyfold, if not orders of magnitude above, block-mining rewards and possible double-spending opportunities.
Furthermore, the boundary between coders and non-coders demarcates an asymmetric-information threshold which poses particular challenges for scaling the DL fintech user base -- e.g. for SCs -- economy wide.
Overall, the consequences for regulation and the design of new financial services -- some of which have not been previously fully anticipated -- fall under the rubric of `boundary effects'.

Fundamentally, then, the scaling of DL-based fintech business models to economy-wide proportions requires both technologists and regulators to confront previously unanticipated consequences of this never-before-witnessed development.
Technologists must face the possibility that the well-tuned internal logic of distributed-consensus incentives can be upset by high-powered external incentives brought to bear upon individual DL transactions.
In turn regulators must face the possibility that traditional-economy working assumptions -- such as `lower TCs are always preferable' -- may not apply straightforwardly to all fintech implementations.
In short, there is a need to develop a cyber-aware profile of concerns and priorities appropriate to the mixed economy featuring cryptocurrency and SCs at scale in parallel with conventional (incumbent) fiat currency and legal contracting.

\section{TCs, risks, information impactedness, and market failure}\label{sec:two}
DL-based business models appear to afford drastic TC-reduction advantages.
Yet it would be erroneous to think that they economize on \emph{every} category of TC. To a degree, this question turns on the definition of TC employed.
Within a neoclassical economics framework, TCs are often portrayed in narrow terms: in currency transactions for instance, as direct trading costs, such as the sum of applicable commissions and the difference between the mid-trading price and the relevant bid- or ask-price.
These are \emph{direct} TCs.
Yet \emph{indirect} TCs form an altogether larger, more extensive and consequential class.

A broader conception of TCs has been developed in the economics of organization and the economics of transaction- and contract-type choice, collectively referred to as Transaction Cost Economics (TCE).\footnote{The field has been recognized with two nobel prizes: Ronald H. Coase in 1991; Oliver E. Williamson in 2009.}
The broader conception of TCs includes transaction-specific
(i) information acquisition and processing costs,
(ii) bargaining and contracting costs,
(iii) monitoring and enforcement costs, and
(iv) the costs of bearing, hedging, or mitigating transaction-specific \emph{risks}.
The last category (iv) has been studied in the finance literature,\footnote{albeit under a different label} where risks impinging upon a proposed trade either reduce the net expected profit or raise the risk-adjusted-return requirement, with the effect of rendering many \emph{potential} trades unprofitable, which thereby are not implemented.

As a result of direct and indirect TCs, particular classes of financial transactions may be undertaken in lower volume than in their absence.
Potentially, the volume may be driven to zero.
If the class of financial transactions under consideration constitutes the entire market -- i.e. under a narrow market definition -- this is referred to as `market failure'.

Individual business models within a market may also find that their transaction volume is diminished -- potentially, even to zero -- as a result of TCs (i)--(iv).
Particularly germane is the concept of \emph{information impactedness}, which is a derivative condition resulting from asymmetric information defined as
\begin{quote}
a situation in which either (1) information is asymmetrically distributed between contracting parties and can be equalized only at great cost, or (2) it is costly to apprise an arbiter of the true information condition should a dispute arise between parties who have identical knowledge of the underlying circumstances \citep[p. 65]{williamson:96}.
\end{quote}
The information asymmetry between programmers and non-pro-grammers can potentially trigger type (1) information impactedness.
In turn, the difficulty of unambiguously apprising a court of (implicit) legal obligations assumed when concluding a Smart (software-only) Contract can potentially trigger type (2) information impactedness.
Indeed for SCs that exist solely as software code with no explicit or implicit legal obligations implied, the cost of apprising an arbiter to resolve a dispute can approach infinity.
Under information impactedness, individuals and firms lean away from entering into a transaction that otherwise would be perceived as mutually beneficial.
When information impactedness is sufficiently severe, customer volume drops to zero, which is the firm-level equivalent to market failure.

The concept of information impactedness captures the consequences of strategic self-serving action that is concealed by asymmetric information.
These consequences include hesitation to transact with the counterparty.
This hesitation may be so strong as to preclude the transaction entirely.
Henceforth we will also refer to the coincidence of asymmetric information and the possibility of self-serving strategic action as giving rise to a costly \emph{risk}, which falls within type (iv) TCs.
We now consider two classes of such risks that are prominent in fintech business models.

\section{Risks}\label{sec:three}
\subsection{Code risks}
Business models based on SCs substitute software code and software execution for legal contracts and legal contract execution.
``Smart contracts are \emph{automation}, not law'' \citep{hazardetal:16}.
This diminishes contemporary legal `contract-dispute risk', but does so by making legal recourse difficult and costly, if not impossible, even for wronged parties.
Instead, SC-based business models are subject to cybersecurity risks and software-code risks.

Instead of Business Email Compromise (BEC) (e.g. for invoice fraud) that off-line bricks-and-mortar firms are subject to, non-DL fintech firms are subject to the risk of direct cyberattacks for exfiltration of personal data and password credentials, for denial of service, and for other nefarious ends.
DL-based fintech firms build upon the culmination of decades of cryptography and security research, and while SHA-256 encryption is currently considered suitably strong, there still remain some areas of vulnerability (e.g. theft of cryptocurrency, tracing coin history through the DL to resolve pseudonyms into identities, Sybil attacks, packet sniffing, and denial of service).

Setting zero-day exploits aside, there is a risk that even well-tested code contains features that allow the system to be gamed -- to the financial advantage of the well-placed software expert -- in a manner not in keeping with the spirit of its intended use, although perhaps not technically illegal, and perhaps not technically constituting `hacking'.
The troubles experienced by Ethereum's Decentralised Autonomous Organisation (DAO) in 2016 are a prominent example of this form of code risk.

Furthermore, the use of DL technology in the form of SCs enables the creation of highly complex cascading transactions with low direct TCs,\footnote{SC platform designers recognize the need to discourage overconsumption of distributed computation resources. The Ethereum platform implements this by charging a microfee (`gas') for each transaction execution.} and low indirect TCs bounded from above by the opportunity cost of time.
Within physically bound systems there are challenges to have long complex transactional chains.\footnote{Consider the example of a house-selling chains.}
With SC-based systems, the cost to extending a transaction chain does not scale up strongly with the length of the chain.
Once triggered however, transactions can rapidly cascade through the chain.
It is entirely feasible to have thousands of transactions completed within a very short timeframe.\footnote{Limited only by the block-formation rate of the DL.}
The complexity is further increase when considering networks of transactions rather than simple chains.
Such transaction networks would be costly and time-consuming to set up within the existing legacy framework of paper contracts, accountants, and lawyers.
But with SCs, incredibly complex transaction networks could be established by a single individual.
In addition, once the networks were triggered they would rapidly cascade across the economy, potentially allowing `cashing out' into the fiat-currency economy before they can be stopped.
It is the ability of one person to orchestrate such complexity within a SC-enabled financial system that poses a new and potentially significant risk.

The bottom line is that code can be \emph{gamed}, just as bricks-and-mortar accounting and financial architecture can, and has been, gamed.
People -- non-programmers in particular -- fear this possibility, regardless of whether software experts consider this to be a genuine concern or not.
This risk -- real and/or perceived -- enters as a TC that hinders customer adoption and utilization of the associated fintech business models, especially on the part of the non-programmer population.

\subsection{Moral-hazard risks}
The boundary between the DL payments ecology and the fiat-currency payments ecology is bridged by \emph{cryptocurrency exchanges} (CCEs).
The DL payments ecology is governed by the distributed-consensus algorithm \citep{nakamoto:08,wattenhofer:16} and its strong (pseudonymous) security yielding provable fairness and non-mani-pulability.
These properties are not shared by the fiat-currency payments ecology.
\emph{Neither}, however, do the properties apply to all of the operational activities of the CCE itself.
Although the operation of a CCE leaves pseudonymous traces in the DL (unless anonymized through a mixing service), for the rest of its operations,  there is at least a degree of asymmetric information.
Compared to the volume and intrusiveness of regulation that bricks-and-mortar banks are subject to, CCEs are, for the time being, lightly regulated\footnote{Some jurisdictions, such as New York State, have introduced compulsory licensing of digital-currency exchanges or `money transmitters'. The New York State license is known as BitLicense. The response of bitcoin exchanges has been to leave New York \citep{guadamuz/marsden:15,roberts:15-f,defilippi:16-w,williams-grut:17-biuk}.} and largely `black-box' operations.

The extent of asymmetric information between the operators of CCEs and their customers has been illustrated vividly by the information released into the public domain by the various investigations and legal proceedings that have been conducted in the wake of the bankruptcy of \emph{Mt. Gox}, the once-dominant Tokyo-based Bitcoin exchange \citep{adelstein/stucky:16,kaminska:16-fta}.

In CCEs, `code risk' coincides with the possibility for strategic self-serving action occluded by asymmetric information -- i.e., with `moral-hazard risk'.
Is the CCE maintaining large enough (i.e. $100\%$) cryptocurrency and fiat-currency balances?
Is the CCE investing and spending enough to protect customers' cryptocurrency from hackers?
Is the CCE investing enough to protect your personal-ID details from being accessed by those with nefarious motives?
Is the CCE front-running its customers' orders?\footnote{This is not even illegal for CCEs operating in light- or no-regulation jurisdictions.}
Is the CCE strategically manipulating withdrawal limits or exchange-completion turn-around times to complete exchanges?
Is the CCE's management operating in a manner that is not short-squeezing itself?\footnote{Inadvertently pushing the price of cryptocurrency upward by its attempt to cover an existing shortage of cryptocurrency balances on its books, as befell Mt. Gox \citep{kaminska:16-fta}.}
Is the CCE engaging in other forms of manipulation, such as manipulation of the exchange rate between the cryptocurrency and fiat currencies?\footnote{See \citet{kaminska:13-fta,kaminska:16-fta,southurst:14,sirer:14-hd,adelstein/stucky:16}.}

Cryptocurrency exchanges are prime examples of fintech businesses that face this particular combination of risks.
Importantly, so do the (potential) customers of cryptocurrency exchanges.
In choosing between exchanges, the customer must gauge and weigh up the software-code risks and moral-hazard risks of each exchange relative to the risks of other exchanges she might use instead.\footnote{Of the 40 exchanges established in the 3 years leading up to January 2013, 18 had closed by April 2013. Often, customers' balances were wiped out. \citep{moore/christin:13}}
These risks act as a brake, slowing adoption and utilization.
In the extreme case, these risks can cause catastrophic loss of confidence in a particular fintech business, or potentially, even in an entire fintech business-model category.

\section{Mixed economy}\label{sec:four}
The `mixed economy' market configuration is of particular commercial and regulatory interest.
In the mixed-economy market configuration, a large-scale cryptocurrency-based SC ecosystem coexists in parallel with a large-scale fiat-currency ecosystem.

The crypto-monetary incentives at work within the Proof-of-Work (PoW) decentralized-consensus algorithm pertain to
(i) block-completion rewards and
(ii) double-spending opportunities.
For a DL which exists in isolation, the incentive for launching a 51\% attack \citep{nakamoto:08,karame/etal:12,rosenfeld:14,pinzon/rocha:16} or a selfish-mining attack \citep{eyal/sirer:14} is precisely the opportunity to appropriate block-completion rewards and to double spend.
These crypto-monetary proceeds are inherently bounded from above.
Double-spending opportunities are limited to the sum of cryptocurrency transactions that the attacker is in a position to reverse.
However, for a SC DL within a mixed-economy market configuration, the upper bound is manyfold higher.

In a mixed economy, SCs allow the creation of dependencies between the fiat-currency ecosystem and the cryptocurrency ecosystem.
This is pivotal because the fiat-currency ecosystem permits and facilitates the creation of fractionally backed (leveraged) financial structures with high positive returns in `good states' and even higher losses in `bad states'.
Through Credit-Default Swaps (CDSs), bad-state losses may be transformed into gains of equal magnitude -- and such CDSs may also be purchased by a third party, or indeed multiple third parties, not only by the individual or firm who has direct exposure to the bad-state losses.
If the CDSs are purchased with borrowed funds, then the total (albeit highly risky) payoff and return are potentially manyfold multiples of the bad-state loss, which in turn is a manyfold multiple of the upper bound on potential double-spending proceeds on the DL.

If provision of a \emph{necessary input} to bringing about the good state is delegated to a SC, then the total financial incentive brought to bear on interfering with the execution of the SC -- and thus for mounting an attack on the DL -- is given by the sum of net CDS payoffs.
Such a necessary input could be, for instance, timely payment on automatically executing smart insurance contracts as envisioned by Emin G\"{u}n Sirer \citep{underwood:16-cacm}.
Notice that there is no inherent upper limit to the number of CDS contracts, nor to the number of purchasers.
Hence, there is no \emph{inherent} upper bound to the financial incentives motivating DL attacks, aside from practical limits associated with nefarious actors' financial resources and the lines of credit they can access.
Notice that these nefarious actors need not be DL node operators (i.e. miners), and that the multitude of different actors who are in line to lose or profit on a grand scale creates a ready market for `attack as a service'.

\section{Virtually zero transaction costs}\label{sec:five}
A common view among regulators and policy makers is that reductions in TCs unambiguously promote economic efficiency, thereby enhancing the welfare-generating potential of the economy, ceteris paribus.\footnote{Distributional implications should always be considered as well.}
A common trait of many digital systems -- no less the case for cryptocurrency-based systems -- is that TCs fall drastically toward zero with their introduction into the economy \citep{shapiro/varian:98,anderson:08}.
Therefore the introduction of cryptocurrency-based systems can transform the TCs of not only national payments and transactions, but on a more fundamental and far-reaching level, of cross-border international payments and transactions.
Cryptocurrency-based systems are expected to play a large role in enhancing the welfare-generating potential of the \emph{global} economy.

Even though DLs and SCs virtually eliminate TCs, there can be unintended consequences in faithfully replicating the form of a traditional service upon a SC-enabled technological platform.
Here we address precisely the existence of non-obvious risk mitigation features that are present in the incumbent, pre-fintech implementations.

To illustrate this effect without prejudicing specific fintech business models, consider the example of DL- and SC-based lotteries, of which there are numerous in existence, primarily as demonstrators.

Bricks-and-mortar lotteries --- such as state 6/49 lotteries in the US and the national 6/49 lottery in the UK -- have been studied extensively, both empirically and mathematically.
Anthony Krautmann and James Ciecka were the first to study what they labelled the ``Trump Ticket'' in parimutuel lotteries, which consists of buying a ticket on every possible number combination \citep{moffitt/ziemba:16,krautmann/ciecka:93-eej,Matheson:01-eej,thaler/ziemba:88-jep}.
When implementation of a Trump-Ticket strategy succeeds in acquiring and storing securely valid tickets for all possible number combinations, it delivers a \emph{share} of the jackpot with certainty.
The share itself is a random variable, which depends on the number of winning tickets held by the crowd of other lottery participants.

Nevertheless, in the bricks-and-mortar world the viability of a Trump-Ticket strategy is frustrated by the frictions, logistical challenges, and TCs of physically buying up and securely storing physical copies of millions of lottery tickets.
Krautmann and Ciecka describe an early attempt to implement the Trump-Ticket strategy:
\begin{quote}
...an Australian consortium made an agreement with a local grocery chain to buy up as many lottery tickets as possible for the \$27 million lottery on 15 February 1992. According to the [Chicago Tribune, 21 Feb 1992] report, the stores processed the tickets in 10,000 increments until time ran out, at which time the consortium had purchased 2.4 million of the 7.1 million possibilities \citep{krautmann/ciecka:93-eej}.
\end{quote}
Despite involving several lotto-ticket sales points in several stores, this consortium succeeded in purchasing tickets on only 33.8\% of the possible number combinations -- a costly undertaking that resulted in a 66.2\% chance of a total loss of funds committed to the Trump-Ticket strategy.
Summarizing previous work \citep{thaler/ziemba:88-jep,matheson/grote:05} Ken Grote and Victor Matheson conclude that
\begin{quote}
the transaction costs involved in purchasing all possible combinations are too high to make this strategy feasible, creating an effective barrier to purchasing the Trump Ticket \citep{grote/matheson:11}.
\end{quote}
The task of purchasing, transporting, and securely storing $n$ lottery tickets within the available time frame before the lottery draw may be understood in production-process terms.
In industry, `Six Sigma' is one of the highest production-quality standards that firms aspire to.
Six Sigma was developed at Motorola and was famously implemented at General Electric. In this quality standard, a production process is refined with the objective of generating fewer than the six standard-deviations quantity of faulty products -- i.e. fewer than 3.4 defective features per 1,000,000.
In a general analysis of this problem, one considers the trade-off between the fault rate (of loss, damage, or other invalidation of a ticket) and the up-front investment, as well as the positive relationship between fault rate and the production rate (of printing, transporting and storing tickets) \citep{kaivanto:17}.
For present purposes we illustrate the phenomenon with a simplified model in which the valid-ticket rate $v\in[0,1]$ is a decreasing function of the number of tickets purchased, $n$, which is an integer between 0 and $N$, the total number of possible tickets in the lottery.\footnote{In $6/49$ lotteries, $N=\frac{49}{6}\cdot\frac{48}{5}\cdot\frac{47}{4}\cdot\frac{46}{3}\cdot\frac{45}{2}\cdot\frac{44}{1} = 13,\!983,\!816$.}
We formalize the relationship in the simplest possible manner $v(n,N)=1-b\frac{n}{N}, b>0$.
For fixed $N$, it will be possible to suppress this term in the valid-ticket-rate function, which becomes $v(n) = 1-bn$.
Here $b$ represents the proportion of invalid tickets when $n=N$.
Let us simplify the presentation further by assuming that the maximum number of \emph{distinct} tickets in the lottery is $N=1000$, and that the `crowd' has already purchased 1000 (randomly selected, uncorrelated) tickets at \$1 per ticket.
Then we may write our expected winnings from purchasing $n$ tickets as follows
\begin{equation}
E[W_n] = \frac{(1-bn)n(n+1000)}{1001}\left(1-\left(1-\frac{1}{1000}\right)^{1001}\right)\;\;,
\end{equation}
and the associated expected gain as
\begin{equation}
E[G_n] = E[W_n] - n - I
\end{equation}
where $\$n$ is the cost of purchasing the $n$ tickets and $\$I$ is the cost of implementing the logistical exercise of purchasing the $n$ tickets.

In contrast, when the lottery is implemented with a DL there are no frictions or transaction costs to prevent the  Trump-Ticket strategy from being profitable.
It can be implemented with a few clicks, literally, in an instant -- importantly, with zero faults, meaning that all tickets purchased are valid $b=0$, and at zero logistical costs $I=0$.

\begin{figure}[!t]
\begin{center}
\caption{Expected net gains as a function of the number of tickets bet; DL-based lottery (in blue), and bricks-and-mortar lottery (in red).} \label{fig:blockch-bnm-exp-gain}
 \scalebox{1.4}[1.4]{
 \includegraphics{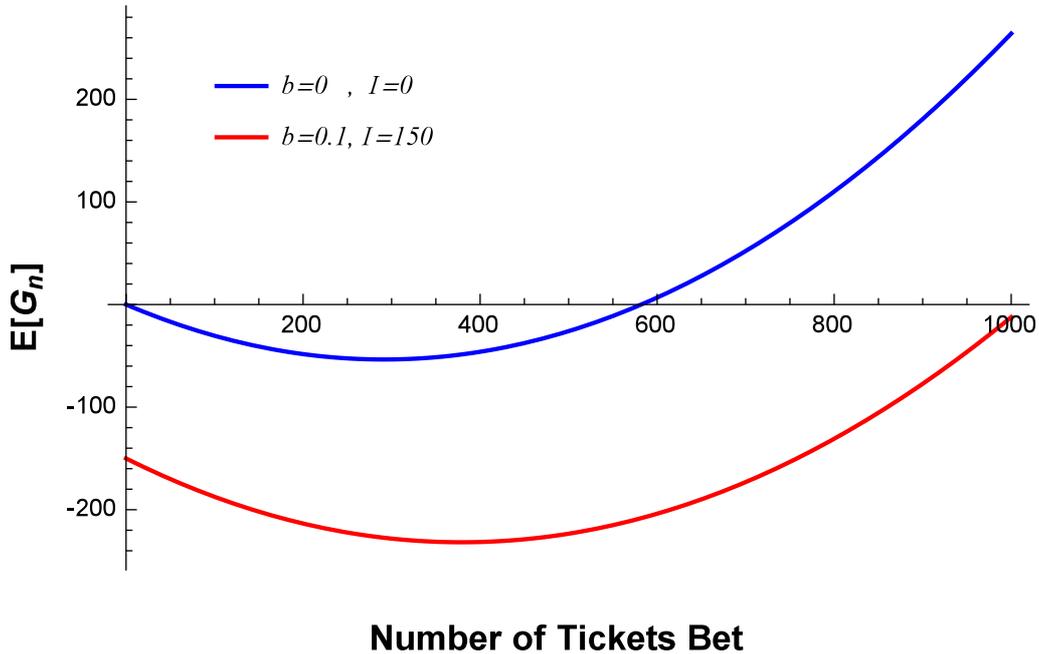}                     
 }
\end{center}
\end{figure}

Figure \ref{fig:blockch-bnm-exp-gain} presents the expected gain associated with each (incomplete) Trump-Ticket strategy $n\leq N = 1000$.
For the bricks-and-mortar lottery -- in which (a) there is a 10\% chance of invalidating a ticket through misprinting or mishandling in transportation or storage, and (b) the cost of underwriting the printing, transportation and storage of tickets is \$150 -- there is no partial or complete Trump-Ticket strategy that yields a positive expected gain (see red curve).
In contrast, any Trump-Ticket strategy purchasing 583 or more tickets on a DL-based lottery secures, in expectation, a positive payoff (see blue curve).

Counterintuitively, the drastic reduction in transaction costs brought about in fintech business models can have unintended consequences. Even something as simple as running a lottery can be a challenge precisely because of the comparative absence of frictions and transaction costs.

\section{Discussion}
Solutions are more readily forthcoming for some of the challenges identified in Sections \ref{sec:three}--\ref{sec:five} than for others.

For problems arising out of virtually zero TCs, simple restrictions on customers (e.g. setting a limit on the number of tickets that may be purchased by each customer) are easily subverted by syndicates of coordinated purchasers in the spirit of `crowdsourcing'.
A more prudent and promising avenue is to side-step the parimutuel format entirely in favor of fixed-odds formats.

In turn, the moral-hazard risks present in CCEs invite a regulatory response.
Re-defining the exchange order issued by the customer to the CCE as a principal-agent contract has in principle the potential to address some of the CCE's moral-hazard risks.
But imagining the customer as `principal' implies that she has sufficient information and expertise to design the principal-agent contract.
Furthermore, principal-agent contracts leave the agent bearing some risk -- which clearly poses a threat to the CCE's solvency and ability to redeem on the exchange contracts the CCE is under obligation to fulfil on behalf of other customers.
Indeed the posited solution to the CCE's moral-hazard risks creates a more complicated set of risks which call for a regulatory response as well.
Either way, a regulatory response is required.
Priority should be alotted to regulation in the first instance, rather than as a patch to deal with additional complications resulting from (as yet unprecedented) principal-agent contracting for exchange services.

What we have called code risk is inherent in DL-based business models, and poses a particular challenge for SC-based business models.
As SC-based business models are developed, two cross-cutting boundaries will take on increasing importance: the boundary between programmers and non-programmers, and the boundary between software-only transactions and software-and-legal-contract-backed transactions.
Trust in code among non-programmers can be built over time, but without enabling legal recourse (i.e. backing each transaction with a legal contract, abandoning software-only execution) an important risk remains.

However an opportunity exists to supplant the incumbent providers of dispute-resolution services, i.e. the legal profession.
Already large-scale online transaction providers such as eBay and Amazon provide their own low-cost dispute-resolution systems.
These dispute-resolution systems are actually the second-line of defense, where the first line is comprised of on-line feedback, user-experience, and reputation metrics systems.
The development of a low-cost and reliable dispute-resolution system as an alternative to legal recourse would go far in dispelling the perceived `code risk' associated with SC-based business models.
This would presumably be paired with a feedback, user-experience, and reputation metrics system for SC developers, if not for individual (standardized) modules within SCs.

Finally, there is the potential for high-powered incentives to cross the boundary between the fiat-currency economy and the cryptocurrency economy as the latter grows in scale to rival the former.
This is the challenge posed by the mixed fiat-and-crypto-currency economy.
And it is a challenge that the developers of decentralized consensus algorithms (DCAs) have not yet addressed head-on.
Nevertheless, it is a high-priority challenge.
Proof-of-Stake (PoS) DCAs are greatly advantaged over PoW DCAs in successfully confronting this challenge.
Prudent financial structuring -- ensuring that the fiat-currency side of an enterprize or of a transaction sequence is not sensitive to timely receipt of payments from the crypto-currency side -- can also pre-empt the emergence of such high-powered incentives for attacking the DL.

\section{Conclusions}
DL-based fintech is distinguished by a \emph{different profile} of risks and transaction costs compared to the incumbent generation of financial services.
Although fintech services' advantages dominate in many categories, there are categories in which fintech services pose greater risks and transaction costs -- even new categories of risks and transaction costs.

Here we have discussed and illustrated code risk, moral-hazard risk, mixed-economy risks, and the unintended consequences of replicating incumbent-generation contract forms within the virtually zero transaction cost environment of fintech.
Already at this early stage in the development of fintech business models it is possible to identify and characterize the types of technical and contractual-form responses with which these challenges may ultimately be resolved.

However, the new profile of risks and transaction costs that we have elaborated here nevertheless highlights the need for a matching and distinctly cyber-aware profile of concerns and priorities -- both among fintech entrepreneurs and fintech consumers, as well as among financial regulators.

\newpage
\bibliography{CryptoLotteries}

